\documentclass[pra,twocolumn,superscriptaddress]{revtex4}
\usepackage[dvipdfmx,hiresbb]{graphicx}
\usepackage{color}
\usepackage{enumerate}
\usepackage{epsfig,subfigure}
\usepackage{amsmath,amssymb,latexsym}
\usepackage{ascmac}
\usepackage{bm}

\newcommand{\ket}[1]{| #1 \rangle}
\newcommand{\ketbra}[2]{| #1 \rangle \langle #2 |}

\begin{document}
\title{Security of quantum key distribution with non-I.I.D. light sources}
\author{Yuichi~Nagamatsu}
\affiliation{Graduate School of Engineering Science, Osaka University,
Toyonaka, Osaka 560-8531, Japan}
\author{Akihiro~Mizutani}
\affiliation{Graduate School of Engineering Science, Osaka University,
Toyonaka, Osaka 560-8531, Japan}
\author{Rikizo~Ikuta}
\affiliation{Graduate School of Engineering Science, Osaka University,
Toyonaka, Osaka 560-8531, Japan}
\author{Takashi~Yamamoto}
\affiliation{Graduate School of Engineering Science, Osaka University,
Toyonaka, Osaka 560-8531, Japan}
\author{Nobuyuki~Imoto}
\affiliation{Graduate School of Engineering Science, Osaka University,
Toyonaka, Osaka 560-8531, Japan}
\author{Kiyoshi Tamaki}
\affiliation{NTT Basic Research Laboratories, NTT Corporation, 3-1, Morinosato Wakamiya Atsugi-Shi, 243-0198, Japan}

\begin{abstract}
Although quantum key distribution (QKD) is theoretically secure, there is a gap between the theory and practice. In fact, real-life QKD may {\it not} be secure because component devices in QKD systems may deviate from the theoretical models assumed in security proofs. To solve this problem, it is necessary to construct the security proof under realistic assumptions on the source and measurement unit. In this paper, we prove the security of a QKD protocol under practical assumptions on the source that accommodate fluctuation of the phase and intensity modulations. 
As long as our assumptions hold, it does not matter at all how the phase and intensity distribute nor whether or not their distributions over different pulses are independently and identically distributed (I.I.D.). Our work shows that practical sources can be safely employed in QKD experiments.
\end{abstract}

\maketitle

Quantum key distribution (QKD) \cite{BB84,review} has attracted much attention in recent years due to its potential to achieve information-theoretically secure communication. QKD enables two distant parties, Alice and Bob, to share a secret key even under interventions by an eavesdropper, Eve. 
So far, many groups have conducted the experiments, including the field demonstrations \cite{demo1, demo2, gisin, demo3, demo4}.
Although QKD is proven to be secure \cite{mayers, LoChau, shorpreskill, koashipreskill, koashiN, kraus, tomamichel1, rrqkd, koashi}, QKD experiments do not perfectly conform to the theoretical models assumed in security proofs due to inevitable imperfections of realistic devices. 
As a result, Eve may exploit the security loopholes that those imperfections open up to learn the distributed key without being detected. 

To solve this problem, it is necessary to generalize the security proof and remove as many demanding assumptions as possible on the component devices. Fortunately, measurement-device-independent (MDI) QKD \cite{mdi1,mdi2,mdi3} closes any security loopholes on the photon-detection side. Moreover, MDIQKD can be implemented with current technologies, and the experimental demonstrations have been reported \cite{mdiexp}. If MDIQKD is widely deployed, the attack must be focused on the source.  Therefore, it is important to prove the security of QKD under realistic assumptions on the source in order to prevent Eve's hacking.

Some previous works analyzed the security based on the source model that accommodates phase modulation (PM) error, which is a commonly faced experimental problem, and showed that the key rate decreases very rapidly with the slight state preparation flaw and with the increase of the channel losses \cite{mdi2,gllp, lopreskill}. This is because the PM error could lead to the leakage of basis information to Eve, which may allow her to perform basis-dependent attacks. Recently, the problem was solved by the loss-tolerant protocol \cite{loss}, which employs the basis mismatched events in estimating Eve's information and shows that the key rate is almost independent of the slight PM error. The finite-key security analysis against general attacks \cite{mizutani} and the experimental demonstrations \cite{feihu, lossmdi} have been reported. The requirements in \cite{loss} are that the single-photon part of each sending state lives in the same qubit space (qubit assumption) and the PM error is assumed to be independently and identically distributed (I.I.D.) over the different pulses. In addition, the sending single-photon states are assumed to be exactly known, i.e., the source is perfectly characterized. In practice, however, due to inevitable imperfections of experimental devices, the requirements are not always fulfilled. 
 For example, the distribution of the PM error may gradually be drifting in time during the experiment. Furthermore, the distributions over the different pulses may be correlated with each other. In addition, it is difficult to precisely estimate the sending states with a  finite sample size. 
Therefore, it is crucial from practical viewpoints to analyze the security of a QKD protocol under these realistic situations.

In this paper, we solve these issues by proving the security of QKD under realistic assumptions on the source, which relaxes assumptions in \cite{loss}. 
Our work is based on the three-state QKD protocol with the qubit assumption as well as the assumption that Alice knows the range in which the phase of each prepared state lies  with a certain probability. As long as our assumptions hold, it does not matter at all how the phase distributes nor whether or not the distributions over different pulses are I.I.D. (non-I.I.D. PM error). 
In addition, we also take into consideration non-I.I.D. amplitude modulation (AM) error. 
The effect of the AM error on the decoy-state method \cite{decoy} was studied in the recent work of \cite{mizutani}, and we combine the idea in \cite{mizutani} with our proof. 
It is worth mentioning that the parameters of the source needed in our proof can be estimated in practical experiments. 
We also perform simulations of the key generation rate to evaluate the performance of a realistic fiber-based QKD system based on our security proof. The results show that the secure keys can be distributed over 100 km even under simultaneous fluctuations of the  phase ($\pm 5^{\circ}$) and the intensity ($\pm5\%$), 
 which indicates  
the feasibility of QKD over long distances with practical light sources. Therefore, our work is an important step towards a secure QKD with realistic devices. 

First, we describe the assumptions we make on the devices for the security proof. We employ the prepare-and-measure three-state protocol. We explain the assumptions on Alice's source followed by the ones of Bob's measurement apparatus.

Alice employs the phase encoding scheme and the decoy-state method \cite{decoy} with two types of decoy states \cite{practicaldecoy}. That is, she uses a phase modulator to encode bit and basis information $c\in\mathcal{C}:=\{0_Z, 1_Z, 0_X\}$ on the relative phase between signal and reference pulses.
Then, she uses an amplitude modulator to encode the intensity information $k\in\mathcal{K}:=\{{\rm s}, {\rm d1}, {\rm d2}\}$ on the total intensity of the double pulses, where ${\rm s}$, ${\rm d1}$, and ${\rm d2}$ denote a signal state and two decoy states, respectively. 
Note that our security analysis applies as well to any other coding schemes and to any number of intensity settings. 
In practice, the phase and the intensity fluctuate due to inevitable imperfections of the source. In addition, the intensities of the signal and reference pulses are not always the same. Here, we propose a device model that accommodates these imperfections. Let $\theta^{(i)}\in\{\theta_{0_Z}^{(i)}, \theta_{1_Z}^{(i)}, \theta_{0_X}^{(i)}\}$, $\mu^{(i)}\in\{\mu_{\rm s}^{(i)}, \mu_{\rm d1}^{(i)}, \mu_{\rm d2}^{(i)}\}$, and $\gamma^{(i)}(>0)$ denote the $i$-th phase, intensity, and ratio of the intensity of the signal pulse to the one of the reference pulse, respectively
($i=1,\dots,N$). The assumptions on the source are as follows.

\noindent
{\bf A-1}. Alice employs a perfectly phase-randomized coherent light and the photon number of each pulse follows a Poisson distribution.

\noindent
{\bf A-2}. Alice selects the setting $c\in\mathcal{C}$ and $k\in\mathcal{K}$ independently at random.

\noindent
{\bf A-3}. For all $c\in\mathcal{C}$, Alice knows the range $[\theta_{c}^{-},\theta_{c}^{+}]$ which satisfies 
${\rm Pr}[\theta_{c}^{-}\leq \theta_{c}^{(i)}\leq\theta_{c}^{+} ]\geq 1-\delta_{\theta}$ for all $i$, and $[\theta_{0_Z}^{-},\theta_{0_Z}^{+}]$, $[\theta_{1_Z}^{-},\theta_{1_Z}^{+}]$, and $[\theta_{0_X}^{-},\theta_{0_X}^{+}]$ do not overlap each other.

\noindent
{\bf A-4}. For all $k\in\mathcal{K}$, Alice knows the range 
$[\mu_{k}^{-}, \mu_{k}^{+}]$ which satisfies
${\rm Pr}[\mu_{k}^{-}\leq \mu_k^{(i)}\leq \mu_{k}^{+}]\geq 1-\delta_{\mu}$ for all $i$, and $1\geq\mu_{\rm s}^{+}$, $\mu_{\rm d1}^{-}>\mu_{\rm d2}^{+}\geq 0$, and $\mu_{\rm s}^{-}>\mu_{\rm d1}^{+}+\mu_{\rm d2}^{-}$ are fulfilled.

\noindent
{\bf A-5}. The mode of the pulse is independent of the choice of the setting $c\in\mathcal{C}$ and $k\in\mathcal{K}$.

\noindent
{\bf A-6}. There are no side-channels.

Satisfying these assumptions, Alice's sending state is expressed as
\begin{widetext}
\begin{align}
\label{sourcemodel}
\hat{\rho}:=\left< \bigotimes_{i=1}^{N}\left\{ \sum_{c\in\mathcal{C}}\sum_{k\in\mathcal{K}}p_{c}p_{k}\hat{P}\left(\left|e^{i\xi}\sqrt{\frac{\mu_{k}^{(i)}}{1+\gamma^{(i)}}}\right>_{r} \left|e^{i(\xi+\theta_{c}^{(i)})}\sqrt{\frac{\gamma^{(i)}\mu_{k}^{(i)}}{1+\gamma^{(i)}}}\right>_s\right)\right\} \right>,
\end{align}
\end{widetext}
where $\left<\ \right>$ represents an integration with weight of a joint probability function of $\theta_{c}^{(i)}$, $\mu_{k}^{(i)}$, $\gamma^{(i)}$ for all $i$, $c\in\mathcal{C}$, and $k\in\mathcal{K}$. This integration is taken over all possible variables of $\theta_{c}^{(i)}$, $\mu_{k}^{(i)}$, and $\gamma^{(i)}$. Here, $p_c$ ($p_k$) is the probability that Alice selects the setting $c\in\mathcal{C}$ ($k\in\mathcal{K}$), $\xi\in[0, 2\pi)$ is the random phase, $\ket{e^{i\xi}\sqrt{\alpha}}$ is a coherent state with intensity (mean photon number) $\alpha$, the subscripts $r$ and $s$ are used to denote the reference and signal modes, respectively, and $\hat{P}(\ket{\cdot}):=\ketbra{\cdot}{\cdot}$. 
Note that Eq. (\ref{sourcemodel}) covers a situation that $\theta_{c}^{(i)}$, $\mu_{k}^{(i)}$, $\gamma^{(i)}$ for all $i$, $c\in\mathcal{C}$, and $k\in\mathcal{K}$ have a classical correlation, and that it does not matter how they distribute nor whether their distributions over different trails are I.I.D. or not.

As for Bob, we make the following assumptions on the measurement  apparatus.

\noindent
{\bf B-1}. Bob chooses a measurement basis at random from the $Z$ and $X$ bases.

\noindent
{\bf B-2}. The detection efficiency is independent of his basis choice.

\noindent
{\bf B-3}. There are no side-channels.

Let $\{\hat{M}_{0_{W'}}, \hat{M}_{1_{W'}}, \hat{M}_{\O_{W'}}\}$ denote Bob's positive operator valued measure (POVM) associated with his basis choice $W'\in\{Z, X\}$, where $\hat{M}_{0_{W'}}$ ($\hat{M}_{1_{W'}}$) and $\hat{M}_{\O_{W'}}$ correspond to the bit value $0$ ($1$) and to an inconclusive outcome, respectively. The POVM operators satisfy  $\hat{M}_{0_{W'}}+\hat{M}_{1_{W'}}+\hat{M}_{\O_{W'}}=\hat{I_d}$, where $\hat{I_d}$ is the identity operator. From B-2, $\hat{M}_{\O_{Z}}=\hat{M}_{\O_{X}}$ is fulfilled \cite{koashiN}.

Next, we describe our three-state protocol. The secret key is extracted from the events where both Alice and Bob have chosen the $Z$ basis in the signal settings. The protocol runs as follows.

{\it 0. Source characterization.} Before starting the protocol, Alice characterizes her source to identify $\theta_{c}^{\pm}$ for all $c\in\mathcal{C}$, $\delta_{\theta}$, $\mu_{k}^{\pm}$ for all $k\in\mathcal{K}$, and $\delta_{\mu}$.

{\it 1. Preparation.} For each pulse, Alice randomly chooses the basis $W^{(i)}\in\{Z, X\}$ with probabilities $p_{Z}$ and $p_X=1-p_{Z}$, respectively, and the intensity setting $k^{(i)}\in\mathcal{K}=\{{\rm s}, {\rm d1}, {\rm d2}\}$ with probabilities $p_{\rm s}$, $p_{\rm d1}$, and $p_{\rm d2}:=1-p_{\rm s}-p_{\rm d1}$, respectively. Then, she selects a bit value $a^{(i)}\in\{0, 1\}$ uniformly at random when $W^{(i)}=Z$ while $a^{(i)}=0$ when $W^{(i)}=X$, i.e., $p_{0_Z}=p_{1_Z}=p_{Z}/2$ and $p_{0_X}=p_{X}$. 
Finally, she prepares the pulses in the state based on the chosen settings and sends it to Bob over a quantum channel. 

{\it 2. Measurement.} For each incoming pulse, Bob randomly chooses the measurement basis $W'^{(i)}\in\{Z, X\}$. Then he performs a measurement in the $W'^{(i)}$ basis and records the outcome $b^{(i)}\in\{0,1,\O\}$. In practice, the measurement device is usually implemented with two single-photon detectors. In this case, the measurement has four possible outcomes $\{0, 1, \O, \perp \}$, which correspond to the bit values $0$, $1$, no detection, and double detection, respectively. For the first three outcomes, Bob assigns what he observes to $b^{(i)}$, and for the last outcome $\perp$ he assigns a random bit value to $b^{(i)}$. 

{\it 3. Basis reconciliation.} After repeating the steps {\it 1-2} $N$ times, Alice announces $\theta_{c}^{\pm}$ for all $c\in\mathcal{C}$, $\delta_{\theta}$, $\mu_{k}^{\pm}$ for all $k\in\mathcal{K}$, $\delta_{\mu}$, her basis choices, and intensity settings while Bob announces his basis choices and those instances with the detection events, over an authenticated public channel. Then, they also announce their bit values for the events where Bob has chosen the $X$ basis. With these information, they identify the following sets:
$\mathcal{S}_{Z,Z,k}:=\{i|W^{(i)}=W'^{(i)}=Z\land k^{(i)}=k\}$ for all $k\in\mathcal{K}$, $\mathcal{S}_{Z,Z,k,{\rm det}}:=\{i|W^{(i)}=W'^{(i)}=Z\land k^{(i)}=k\land b^{(i)}\neq\O\}$ for all $k\in\mathcal{K}$, $\mathcal{S}_{c, X, k}:=\{i|a^{(i)}_{W^{(i)}}=c\land W'^{(i)}=X\land k^{(i)}=k\}$ for all $c\in\mathcal{C}$ and $k\in\mathcal{K}$, and $\mathcal{S}_{c, b_X, k}:=\{i|a^{(i)}_{W^{(i)}}=c\land W'^{(i)}=X\land k^{(i)}=k\land b^{(i)}=b\}$ for all $c\in\mathcal{C}$, $k\in\mathcal{K}$, and $b\in\{0,1\}$.

{\it 4. Generation of sifted key and parameter estimation.} 
A sifted key pair (${\bf Z}_{A}$, ${\bf Z}_{B}$) is generated by choosing a random sample of $\mathcal{S}_{Z,Z,{\rm s},{\rm det}}$, while the rest of $\mathcal{S}_{Z,Z,{\rm s},{\rm det}}$ is used to estimate the bit error rate $e_Z$. Then, they calculate the gain in the $Z$ basis $Q_{{\rm det}|Z,Z,k}:=|\mathcal{S}_{Z,Z,k,{\rm det}}|/|\mathcal{S}_{Z,Z,k}|$ for all $k\in\mathcal{K}$, 
which is the fraction of the number of the detection events given the setting $\{W=W'=Z\land k\in\mathcal{K}\}$. Here, $|*|$ denotes the length of the set $*$. 
Next, they compute $Q_{b|c,X,k}:=|\mathcal{S}_{c, b_X, k}|/|\mathcal{S}_{c, X, k}|$ for all $c\in\mathcal{C}$, $k\in\mathcal{K}$, and $b\in\{0,1\}$, which is the fraction of the number of the 
events where Bob has obtained $b$ given the setting $\{c\in\mathcal{C}\land W'=X \land k\in\mathcal{K}\}$. With these experimentally available data, they use the decoy-state method to calculate the single-photon contribution to $Q_{{\rm det}|Z,Z,k}$ and $Q_{b|c,X,k}$. Finally, by using the single-photon part of the data, they calculate the upper bound on the phase error rate of single-photon states [Eq. (\ref{EX1})] in (${\bf Z}_{A}$, ${\bf Z}_{B}$).

{\it 5. Postprocessing.} Alice and Bob perform error correction followed by privacy amplification on (${\bf Z}_{A}$, ${\bf Z}_{B}$) to generate a secret key. 

Now, we show the protocol is secure against coherent attacks. In order to generate a secret key from the sifted bits, Alice and Bob have to estimate the min-entropy of the single-photon part, which quantifies the amount of information leaked to Eve \cite{kraus, tomamichel1}. For this, we employ the phase error rate of the single-photon part \cite{koashiN, koashi} that is related to the min-entropy. Given the proper estimation of the rate, we can generate a secret key by performing privacy amplification based on the estimated phase error rate. Note that for simplicity, we study the asymptotic scenario, where the number of emitted pulses $N$ is infinite.

For the estimation of the phase error rate, we consider the worst case, in which sending states are regarded as the so-called {\it tagged} states unless both  $\theta_{c}^{-}\leq\theta_{c}^{(i)}\leq\theta_{c}^{+}$ and  $\mu_{k}^{-}\leq\mu_{k}^{(i)}\leq\mu_{k}^{+}$ are simultaneously satisfied.  
The tag tells Eve the bit and basis information of the state, and therefore she can learn perfect information on those instances without causing any errors. It is convenient to introduce a probability $p_{\rm t}:=1-(1-\delta_{\theta})(1-\delta_{\mu})$, which is the probability of the sending state being tagged.  
We pessimistically assume that the binary entropy of the phase error rate of the tagged single-photon states is $1$ due to the worst case scenario.
Therefore, the secret key generation rate $R$ (per pulse) is lower bounded by
\begin{multline}
\label{keyrate}
R\geq p_Z^2 p_{\rm s}\{ Q_{{\rm u},1, {\rm det}|Z,Z,{\rm s}}\left(1-h(e_{X,{\rm s},{\rm u},1})\right) \\ -Q_{{\rm det}|Z,Z,{\rm s}}f_{\rm EC}h(e_Z)\},
\end{multline}
whenever the right-hand side is positive \cite{gllp}. Here, $Q_{{\rm u},1, {\rm det}|Z,Z,{\rm s}}$ is a fraction of $Q_{{\rm det}|Z,Z,{\rm s}}$ that originates from the untagged single-photon states, $e_{X,{\rm s},{\rm u},1}$ is the phase error rate of the untagged single-photon states in the signal settings, $f_{\rm EC}$ is the efficiency of the error correcting code, and $h(x)$ is the binary entropy function. 
In the following, we give the definition of $e_{X,{\rm s},{\rm u},1}$. Then, we describe the way to estimate the upper bound on $e_{X,{\rm s},{\rm u},1}$ from the experimentally available data in the protocol.

For simplicity, we first consider the case where $\gamma^{(i)}=1$ for all $i$, i.e., the intensities of double pulses are always the same. The general case where $\gamma^{(i)}\neq1$ is described in Appendix A. 
According to Eq. (\ref{sourcemodel}), the untagged single-photon part prepared in the signal settings is expressed as
\begin{equation}
\label{singlepart}
\hat{\rho}_{{\rm s},{\rm u}, 1}=\left<\bigotimes_{i=1}^{N_{{\rm s},{\rm u}, 1}}\sum_{c\in\mathcal{C}}p_{c}\hat{P}(\ket{\theta_{c}^{(i)}}) \right>,
\end{equation}
where $N_{{\rm s},{\rm u},1}$ is the number of events where Alice has  emitted untagged single-photon states in the signal settings, and the interval of integration with respect to $\theta_{c}^{(i)}$ is $[\theta_{c}^{-}, \theta_{c}^{+}]$. 
Here, we define a qubit state as  $\ket{\theta}:=(\ket{1}_{r}\ket{0}_{s}+e^{i\theta}\ket{0}_{r}\ket{1}_{s})/\sqrt{2}$, where $\ket{n}_{r (s)}$ denotes $n$-photon state of reference (signal) pulse, and the set of two pure states $\{\ket{1}_{r}\ket{0}_{s},\ket{0}_{r}\ket{1}_{s}\}$ forms a qubit basis. In what follows, the eigenstates in the $Y$ basis are defined as $\ket{0_Y}:=\ket{1}_{r}\ket{0}_{s}$ and $\ket{1_Y}:=\ket{0}_{r}\ket{1}_{s}$, and the ones in the $Z$ ($X$) basis are $\ket{0_Z}:=(\ket{0_Y}+\ket{1_Y})/\sqrt{2}$ and $\ket{1_Z}:=(-i\ket{0_Y}+i\ket{1_Y})/\sqrt{2}$  ($\ket{a_X}:=(\ket{0_Z}+(-1)^{a}\ket{1_Z})/\sqrt{2}$ with $a\in\{0,1\}$). 

The emission of the $Z$- ($X$-) basis states can be equivalently described as follows. Alice first prepares $\ket{\psi_Z^{(i)}}_{AB}=\frac{1}{\sqrt{2}}\sum_{a=0,1}\ket{a_Z}_A\ket{\theta_{a_Z}^{(i)}}_B$ ($\ket{\psi_X^{(i)}}_{AB}=\ket{0_X}_A\ket{\theta_{0_X}^{(i)}}$). Then, she measures the system $A$ in the $Z$ ($X$) basis before sending Bob only system $B$. 
In order to define the phase error rate, we convert the actual protocol into the virtual one in such a way that it is equivalent from Eve's perspective. In the virtual protocol (we refer Appendix A of \cite{loss} and Appendix D of \cite{mizutani} for detailed explanations), Alice and Bob measure systems $A$ and $B$ in the $X$ basis (complementary basis to the $Z$ basis) given the preparation of $\ket{\psi_Z^{(i)}}_{AB}$. 
In these virtual events, Alice sends out $\ket{\theta_{a_{\rm vir}}^{(i)}}_{B}$ with a probability $p_{a_{\rm vir}}^{(i)}$, which are respectively written as
\begin{align} 
\label{VirState}
\hat{P}(\ket{\theta_{a_{\rm vir}}^{(i)}})&={\rm Tr}_{A}\left[\hat{P}(\ket{a_X}_A)\hat{P}(\ket{\psi_{Z}^{(i)}}_{AB})\right]/p_{a_{\rm vir}}^{(i)} \notag \\ 
&=\begin{cases}
\hat{P}(\ket{(\theta_{0_Z}^{(i)}+\theta_{1_Z}^{(i)})/2})& (\text{$a=0$})\\
\hat{P}(\ket{(\theta_{0_Z}^{(i)}+\theta_{1_Z}^{(i)})/2+\pi})& (\text{$a=1$}),
\end{cases}
\end{align}
and
\begin{align} 
\label{VirPro}
p_{a_{\rm vir}}^{(i)}&={\rm Tr}\left[\hat{P}(\ket{a_X}_A)\hat{P}(\ket{\psi_{Z}^{(i)}}_{AB})\right] \notag \\ &= \frac{1}{2}\left( 1+(-1)^{a}\cos{\left( \frac{\theta_{0_Z}^{(i)}-\theta_{1_Z}^{(i)}}{2} \right)} \right),
\end{align}
where ${\rm Tr}_{A}$ denotes the partial trace over system $A$. In the virtual protocol, Alice first prepares 
\begin{equation}
\label{virtualstate}
\hat{\rho}_{AB}=\left<\bigotimes_{i=1}^{N_{{\rm s},{\rm u}, 1}}\hat{P}\left(\sum_{j=1}^{5}\sqrt{p_{j}^{(i)}}\ket{j}_A\ket{\psi_{j}^{(i)}}_B\right)\right>, 
\end{equation}
where $p_{1}^{(i)}=p_{Z}^{2}p_{0_{\rm vir}}^{(i)}$, $p_{2}^{(i)}=p_{Z}^2p_{1_{\rm vir}}^{(i)}$, $p_{3}^{(i)}=p_{4}^{(i)}=p_{Z}p_{X}/2$, $p_{5}^{(i)}=p_X$, $\ket{\psi_{1}^{(i)}}=\ket{\theta_{0_{\rm vir}}^{(i)}}$, $\ket{\psi_{2}^{(i)}}=\ket{\theta_{1_{\rm vir}}^{(i)}}$, $\ket{\psi_{3}^{(i)}}=\ket{\theta_{0_Z}^{(i)}}$, $\ket{\psi_{4}^{(i)}}=\ket{\theta_{1_Z}^{(i)}}$, and $\ket{\psi_{5}^{(i)}}=\ket{\theta_{0_X}^{(i)}}$. Then, she sends system $B$ to Bob before they perform a collective measurement characterized by the POVM 
$\hat{F}_{j',b}$ with $j'\in\{1,2,\dots,6\}$ and $b\in\{0,1,\O\}$, where $\hat{F}_{j',b}=\hat{P}(\ket{j'}_{A})\otimes\hat{M}_{b_X,B}$ for $j'\in\{1,2,3,4\}$, $\hat{F}_{5,b}=\hat{P}(\ket{5}_{A})\otimes p_{X}\hat{M}_{b_X,B}$, and $\hat{F}_{6,b}=\hat{P}(\ket{5}_{A})\otimes p_{Z}\hat{M}_{b_Z,B}$. 
The reason why we can introduce $\hat{\rho}_{AB}$ in Eq. (\ref{virtualstate}) is that Eve's accessible quantum information to the sending system $B$ is the same as the one in the actual protocol, i.e., ${\rm Tr}_{A}[\hat{\rho}_{AB}]=\hat{\rho}_{{\rm s},{\rm u}, 1}$ is satisfied. 
From this, we can compute the probability of obtaining $j'$ and $b$ in the $i$-th trial, which depends on the previous outcomes. Thus, by taking the summation of such probabilities over $i$ ($i=1,\dots,N_{{\rm s},{\rm u},1}$), Azuma's inequality \cite{azuma} gives the actual occurrence number of such events \cite{loss, mizutani, azuma1, azuma2}. Consequently, the fraction of the number of events where Alice and Bob obtain $a$ and $b$, respectively, given that Alice sends untagged virtual state ($j'$=1 or 2), $Y_{a,b|{\rm vir, u}}$, and the fraction of the number of events where Bob obtains $b$ given that Alice sends untagged single-photon states under the setting $\{c\in\mathcal{C}\land W'=X\}$, $Y_{b|c,X,{\rm u},1}$,  
are respectively given by
\begin{equation}
\label{azuma1}
Y_{a,b|{\rm vir},{\rm u}}=\frac{1}{N_{{\rm s},{\rm u},1}}\sum_{i=1}^{N_{{\rm s},{\rm u},1}}\left< p_{a_{\rm vir}}^{(i)}{\rm Tr}\left[\hat{D}_{b_X}^{(i)}
\hat{P}(\ket{\theta_{a_{\rm vir}}^{(i)}}) \right] \right>,
\end{equation}
and 
\begin{equation}
\label{azuma2}
Y_{b|c,X,{\rm u},1}=\frac{1}{N_{{\rm s},{\rm u},1}}\sum_{i=1}^{N_{{\rm s},{\rm u},1}}\left<{\rm Tr}\left[\hat{D}_{b_X}^{(i)}\hat{P}(\ket{\theta_{c}^{(i)}}) \right]\right>,
\end{equation}
where $\hat{D}_{b_X}^{(i)}$ is an arbitrary operator 
including Eve's operation and Bob's $i$-th $X$-basis measurement with the outcome $b$.

Thus, $e_{X,{\rm s},{\rm u},1}$ is defined as
\begin{align}
\label{ex}
e_{X,{\rm s},{\rm u},1}:=&\frac{Y_{0,1|{\rm vir},{\rm u}}+Y_{1,0|{\rm vir},{\rm u}}}{Y_{0,0|{\rm vir},{\rm u}}+Y_{0,1|{\rm vir},{\rm u}}+Y_{1,0|{\rm vir},{\rm u}}+Y_{1,1|{\rm vir},{\rm u}}} \notag \\
=&\frac{Y_{0,1|{\rm vir},{\rm u}}+Y_{1,0|{\rm vir},{\rm u}}}{Y_{{\rm det}|Z,Z,{\rm u},1}},
\end{align}
where $Y_{{\rm det}|Z,Z,{\rm u},1}$ 
is the yield of untagged single-photon states in the $Z$ basis, i.e., the fraction of the detection events given that Alice has sent out untagged single-photon states under
the setting $\{W=W'=Z\}$. Note that $Y_{a,b|{\rm vir},{\rm u}}$, $Y_{b|c,X,{\rm u},1}$, and $Y_{{\rm det}|Z,Z,{\rm u},1}$ do not depend on the intensity setting $k\in\mathcal{K}$ because Eve cannot distinguish the decoy states from the signal states and the only information available to Eve is the number of photons in a sending state \cite{decoy}.
In the second equality of Eq. (\ref{ex}), we have used the assumption B-2 in Sec. I\hspace{-.1em}I.  

In what follows, we provide the way to estimate the upper bound on the phase error rate $e_{X,{\rm s},{\rm u},1}$ from the experimentally available data. Note that the denominator in Eq. (\ref{ex}), $Y_{{\rm det}|Z,Z,{\rm u},1}$, can be lower bounded by the decoy-state method [Eq. (\ref{Ydetzz1})], and as a result, the lower bound $Y_{{\rm det}|Z,Z,{\rm u},1}^{-}$ can be directly calculated by using experimentally available data \cite{mizutani}.
The next step is to estimate the upper bound on $Y_{a,b|{\rm vir,{\rm u}}}$, which is the numerator in Eq. (\ref{ex}). The upper bound is given by 
\begin{align}
\label{shortcal}
Y_{a,b|{\rm vir},{\rm u}}\leq p_{a_{\rm vir}}^{+}\sum_{c\in\mathcal{C}}g_{a, c}^{+}Y_{b|c,X,{\rm u},1},
\end{align}
where $p_{a_{\rm vir}}^{+}$ is the maximum value of $p_{a_{\rm vir}}^{(i)}$ in Eq. (\ref{VirPro}) under the constraint of $\theta_{c}^{-}\leq \theta_{c}^{(i)} \leq \theta_{c}^{+}$. The derivation of Eq. (\ref{shortcal}) and  the definition of $g_{a, c}^{+}$ are described in Appendix C. Note that $p_{a_{\rm vir}}^{+}$ and $g_{a, c}^{+}$ can be calculated.

Finally, we obtain the upper bound on $Y_{a,b|{\rm vir},{\rm u}}$ as follows.
From the decoy state method \cite{mizutani}, we estimate $Y_{b|c,X,{\rm u},1}^{-}$ and  $Y_{b|c,X,{\rm u},1}^{+}$, which are the lower and upper bound on $Y_{b|c,X,{\rm u},1}$, respectively [Eqs. (\ref{Ydetzz1b}), (\ref{YbcX1})]. When $g_{a,c}^{+}$ is positive (negative), we replace $Y_{b|c,X,{\rm u},1}$ in Eq. (\ref{shortcal}) with  $Y_{b|c,X,{\rm u},1}^+$ ($Y_{b|c,X,{\rm u},1}^-$) to obtain the upper bound on $Y_{a,b|{\rm vir},{\rm u}}$, which we denote by $Y_{a,b|{\rm vir},{\rm u}}^{+}$.
With this way, we have the upper bound on the phase error rate as
\begin{equation}
\label{EX1}
e_{X,{\rm s},{\rm u},1}\leq e_{X,{\rm s},{\rm u},1}^{+}=\frac{Y_{0,1|{\rm vir},{\rm u}}^{+}+Y_{1,0|{\rm vir},{\rm u}}^{+}}{Y_{{\rm det}|Z,Z,{\rm u},1}^{-}},
\end{equation}
which can be calculated by using the experimentally available data.

In the following, we present the simulation results for our three-state protocol with non-I.I.D. light sources. We consider two types of simulations. In the first simulation, in order to see the effect of the non-I.I.D. PM error, 
we assume that Alice possesses the single-photon source with the PM error. In the second simulation, she is assumed to use the coherent source with both the non-I.I.D. PM and AM errors.

We begin with the description of the model of the PM error. We assume that $\theta_{c}^{(1)},\ldots,\theta_{c}^{(N)}$ are I.I.D. and follow the same Gaussian distribution which has a mean of  $\bar{\theta}_{c}$ and a standard deviation of $\theta/5.33$, so that the phase fluctuates at most $\pm\theta$, i.e., $[\theta_{c}^{-}, \theta_{c}^{+}]=[\bar{\theta}_{c}-\theta, \bar{\theta}_{c}+\theta]$ and $\delta_{\theta}=10^{-7}$ are satisfied. 
Here, $\bar{\theta}_{0_Z}=0$, $\bar{\theta}_{1_Z}=\pi$, and $\bar{\theta}_{0_X}=\pi/2$. These I.I.D. properties are assumed
only to obtain the parameters $Q_{{\rm det}|Z,Z,k}$ and $Q_{b|c, X, {k}}$, which are measured in the actual experiments.
We emphasize that Alice does not know these I.I.D. properties, and we apply the analysis assuming non-I.I.D. sources. 

The assumed system parameters are as follows. We consider a fiber-based QKD system with an attenuation coefficient $0.2$ dB/km. That is, the channel transmittance is $\eta_{\rm ch}=10^{-0.2l/10}$, where $l$ is the fiber length. Bob uses a measurement setup with two single photon detectors: they have the detection efficiency of $\eta_{\rm B}=0.15$ and the dark count rate of $p_{\rm d}=5\times10^{-7}$. Also, the efficiency of the error correcting code is $f_{\rm EC}=1.22$, and the $Z$ basis (signal settings) is selected with the probability $p_Z=2/3$ ($p_{\rm s}=1/3$). For simplicity, we do not consider the misalignment of the detector at Bob's side.

\begin{figure}[t]
\includegraphics[width=7.5cm]{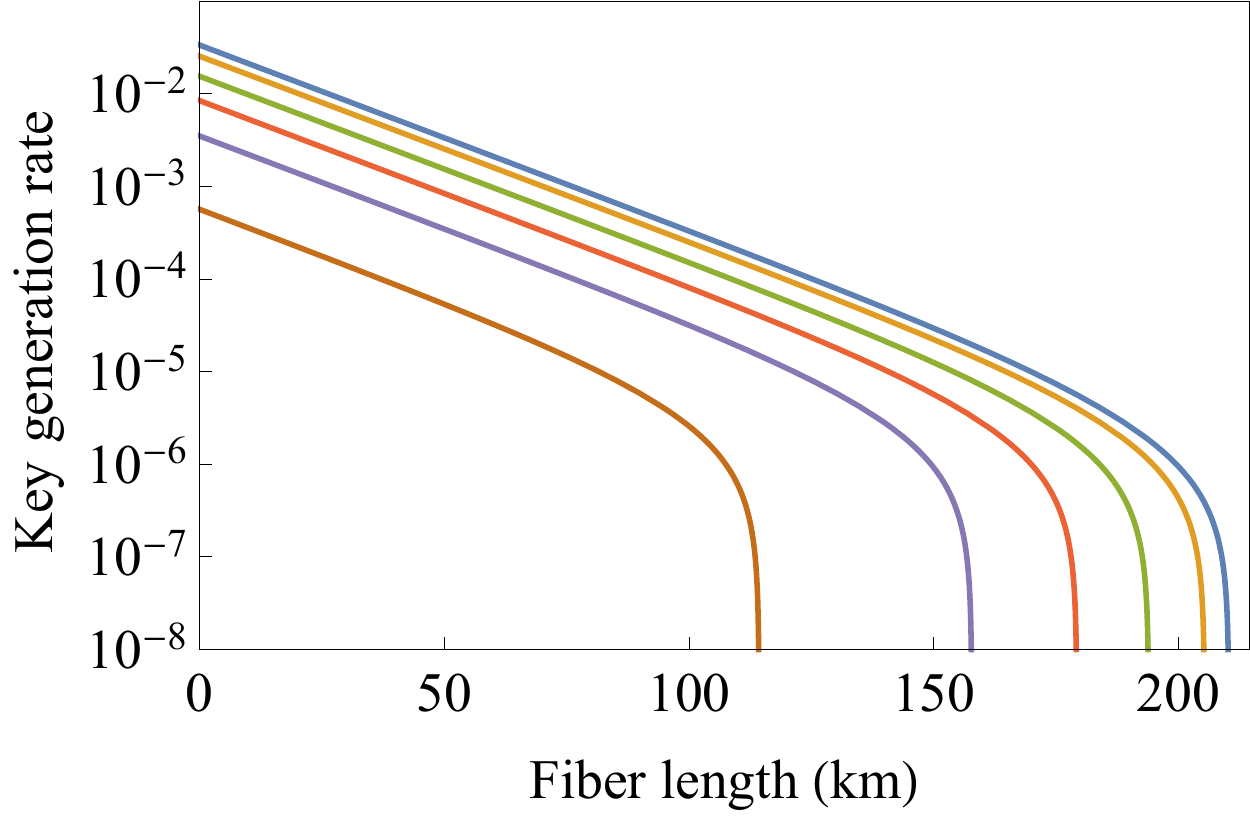}
\caption{(Color online) Secret key generation rate (per pulse) vs fiber length for the first simulation. The key rates are plotted in logarithmic scale for $\theta=0^\circ, 1^\circ, 3^\circ, 5^\circ, 7^\circ, 9^\circ$ (from right to left).}
\label{s1}
\end{figure}

FIG. \ref{s1} shows the resulting secret key generation rate $R$ (per  pulse) as a function of the fiber length for the first simulation (see also Appendix D). 
The key rates are plotted in logarithmic scale for $\theta=0^\circ, 1^\circ, 3^\circ, 5^\circ, 7^\circ, 9^\circ$, which reflects the accuracy of the phase modulator. 

\begin{figure}[b]
\includegraphics[width=7.5cm]{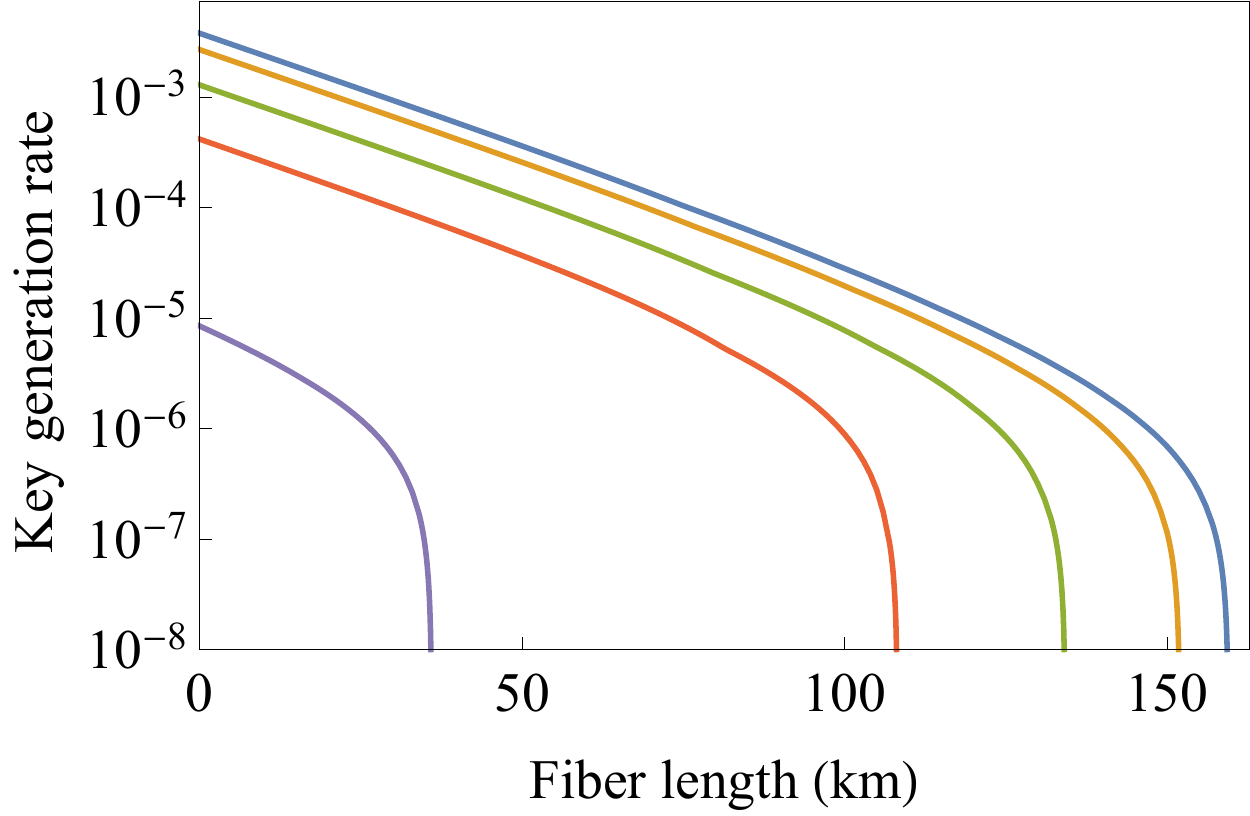}
\caption{(Color online) Secret key generation rate (per pulse) vs fiber length for the second simulation. Numerically optimized secret key rates (in logarithmic scale) are obtained for $(\theta, x)=(0^\circ, 0), (1^\circ, 1), (3^\circ, 3), (5^\circ, 5), (7^\circ, 7)$ (from right to left).}
\label{s2}
\end{figure}

In the second simulation, we assume that $\mu_{k}^{(1)},\ldots,\mu_{k}^{(N)}$ are also I.I.D. and follow the same Gaussian distribution which has a mean of $\bar{\mu}_{k}$ and a standard deviation of $(\mu_{k}^{+}-\bar{\mu}_{k})/5.33$, so that $\delta_{\mu}=10^{-7}$ is satisfied. 
The intensity is assumed to fluctuate at most $x$\%. That is, $\mu_k^{\pm}=(1\pm 0.01x)\bar{\mu}_k$.
For the evaluation, we numerically optimize the key rate $R$ over $\{\bar{\mu}_{\rm s}, \bar{\mu}_{\rm d1}\}$. Note that we fix the intensity of the weakest decoy state to $\bar{\mu}_{\rm d2}=2\times10^{-4}$. FIG. \ref{s2} shows the result for $(\theta, x)=(0^\circ, 0), (1^\circ, 1), (3^\circ, 3), (5^\circ, 5), (7^\circ, 7)$, which reflect the accuracy of the phase and amplitude modulator, respectively (see also Appendix D). 
From these results, we see that the key can still be generated over long distances, which indicates the feasibility of QKD in the presence of high channel loss. Note that the result in FIG. \ref{s2} can be improved with a better analysis of the decoy-state method under intensity fluctuations.

In conclusion, we have proved the security of a QKD protocol under the practical situation where users have limited control over the phase, the intensity, and the ratio of the intensity of the signal pulse to the one of the  reference pulse. As long as our assumptions hold, it does not matter how they  distribute nor whether their distributions are I.I.D. or not. Based on our security analysis, we have made simulations for the key generation rate, which shows a secret key can be securely distributed over long distances with imperfect source devices. We believe that 
our work is an important step to construct a truly secure QKD with realistic devices.

We thank Hoi-Kwong Lo, Marcos Curty, Koji Azuma, Takanori Sugiyama, and Yuki Hatakeyama for helpful discussions.
YN acknowledges support from the JSPS Grant-in-Aid for Scientific Research (A) 25247068 and (B) 15H03704. 
KT acknowledges support from the National Institute of Information and Communications Technology (NICT).

\appendix

\section{The case where the intensities of the signal and reference pulses are different}
In this appendix, we generalize the security proof in the main text to the case where the intensities of the signal and reference pulses are different. That is, the ratio of the intensity of the signal pulse to the one of the reference pulse   $\gamma^{(i)}\neq1$. Here, we show that, even in this case, the upper bound on the phase error rate is the same as the one in the main text [Eq. (\ref{EX1})], where we assume $\gamma^{(i)}=1$.

According to Eq. (\ref{sourcemodel}), given the setting $c\in\mathcal{C}$, the prepared untagged single-photon state is a classical mixture of a qubit state $\ket{\theta_{c}^{(i)},\phi^{(i)}}:=\cos(\phi^{(i)}/2)\ket{0_Y}+e^{i\theta_{c}^{(i)}}\sin(\phi^{(i)}/2)\ket{1_Y}$ 
with $\theta_{c}^{-}\leq \theta_{c}^{(i)}\leq\theta_{c}^{+}$ and $0< \phi^{(i)}<\pi$, where $\phi^{(i)}:=2\tan^{-1}\sqrt{\gamma^{(i)}}$. 
Now, we introduce a filter operation whose successful operation is given by $\hat{F}:=q\hat{P}(\ket{0_{Y}})+(1-q)\hat{P}(\ket{1_{Y}})$, with $0\leq q \leq1$ and $q\neq1/2$. This operation uniformly lifts up all the states on the $X$-$Z$ plane of the Bloch sphere and transforms a state $\hat{\rho}$ with Bloch vector $\{x,0,z\}$ to the state $\hat{F}\hat{\rho}\hat{F}^{\dagger}/p$, whose Bloch vector is $\{f(q)x, (2q-1)/(1-2q+2q^2),f(q)z\}$ with $f(q):=2(1-q)q/(1-2q+2q^2)$. Here, $p:=q^2-q+1/2$ is the success probability of the filtering operation which is the same for all the states on the $X$-$Z$ plane. Applying this filter to $\hat{P}(\ket{\theta_{c}^{(i)}, \pi/2})$, whose Bloch vector is $\{\sin{\theta_{c}^{(i)}}, 0,  \cos{\theta_{c}^{(i)}}\}$, for all $c\in\mathcal{C}$, one obtains $\hat{P}(\ket{\theta_{c}^{(i)}, \phi^{(i)}})=\hat{F}\hat{P}(\ket{\theta_{c}^{(i)}, \pi/2})\hat{F}^{\dagger}/p$ with Bloch vector $\{\sin{\phi^{(i)}}\sin{\theta_{c}^{(i)}}, \cos{\phi^{(i)}}, \sin{\phi^{(i)}}\cos{\theta_{c}^{(i)}}\}$ when the relation of $\tan{\phi^{(i)}}=(2q-2q^2)/(2q-1)$ holds. From this, we can rewrite the transmission rate of the actual states as ${\rm Tr}[\hat{D'}_{b_X}^{(i)}\hat{P}(\ket{\theta_{c}^{(i)}, \pi/2})]$, where  $\hat{D'}_{b_X}^{(i)}:=\hat{F}^{\dagger}\hat{D}_{b_X}^{(i)}
\hat{F}/p$. Importantly, $\hat{D'}_{b_X}^{(i)}$ is independent of $c$. 
This means that even if Alice sends $\ket{\theta_{c}^{(i)},\phi^{(i)}}$ in reality, we are allowed to convert the problem with $\phi^{(i)}\neq\pi/2$ ($\gamma^{(i)}\neq1$) to the case where $\phi^{(i)}=\pi/2$ ($\gamma^{(i)}=1$). This is because the proof with $\gamma^{(i)}=1$ is valid for any operator $\hat{D}_{b_X}^{(i)}$ \cite{loss}.

\section{DECOY-STATE ANALYSIS}
In this appendix, we show how the decoy-state method is used to derive the security bounds presented in the main text, namely, the lower bound on $Y_{{\rm det}|Z,Z,{\rm u},1}$ and the lower and upper bound on $Y_{b|c,X,{\rm u},1}$. The derivation is based on the previous work of \cite{mizutani}, which generalizes the idea of one-signal and two-decoy method \cite{practicaldecoy} to the case where the intensity of each pulse fluctuates and it lies within $[\mu_{k}^{-}, \mu_{k}^{+}]$.

Applying the method of \cite{mizutani} to our work, $Y_{{\rm det}|Z,Z,{\rm u},1}$ can be lower bounded by a function of $Q_{{\rm det}|Z,Z,k,{\rm u}}$ with $k\in\mathcal{K}$, which is the fraction of the number of the detection events given that Alice has sent the untagged states under the setting $\{W=W'=Z\land k\in\mathcal{K}\}$. Since $Q_{{\rm det}|Z,Z,k,{\rm u}}$ cannot be directly obtained in the experiments, we consider to bound it by $Q_{{\rm det}|Z,Z,k}$, which is an  experimentally available data in the protocol. Eve may control the detection efficiency of the tagged states at Bob's side, and therefore, the following condition generally holds:
\begin{equation}
\label{decoy1}
\frac{Q_{{\rm det}|Z,Z,k}-p_{\rm t}}{1-p_{\rm t}}\leq Q_{{\rm det}|Z,Z,k,{\rm u}}\leq \frac{Q_{{\rm det}|Z,Z,k}}{1-p_{\rm t}},
\end{equation}
where the equality holds for the first inequality if and only if Eve makes all the tagged states detected  at Bob's side, while the equality for the second one holds if and only if Eve blocks all the tagged states. By combining the bounds given in \cite{mizutani} and Eq. (\ref{decoy1}), we obtain the lower bound on $Y_{{\rm det}|Z,Z,{\rm u},1}$ as
\begin{align}
\label{Ydetzz1}
&Y_{{\rm det}|Z,Z,{\rm u},1}\geq  Y_{{\rm det}|Z,Z,{\rm u},1}^{-}=\frac{\mu_{\rm s}^{-}}{(\mu_{\rm s}^{-}-\mu_{\rm d1}^{+}-\mu_{\rm d2}^{-})(\mu_{\rm d1}^{+}-\mu_{\rm d2}^{-})} \notag \\
& \times\Big\{ \frac{(Q_{{\rm det}|Z,Z, {\rm d1}}-p_{\rm t})e^{\mu_{\rm d1}^{-}}-Q_{{\rm det}|Z,Z, {\rm d2}}e^{\mu_{\rm d2}^{+}}}{1-p_{\rm t}}\notag \\ &- \frac{(\mu_{\rm d1}^{+})^2-(\mu_{\rm d2}^{-})^2}{(\mu_{\rm s}^{-})^2}\Big(\frac{Q_{{\rm det}|Z,Z, {\rm s}}e^{\mu_{\rm s}^{+}}}{1-p_{\rm t}}-Y_{{\rm det}|Z,Z,{\rm u},0}^{-}\Big) \Big\},
\end{align}
where $Y_{{\rm det}|Z,Z,{\rm u},0}^{-}$ is the lower bound on the yield of untagged vacuum states in the $Z$ basis, which is given by
\begin{align}
\label{Ydetzz0}
&Y_{{\rm det}|Z,Z,{\rm u},0}^{-}=\notag \\ &{\rm max}\Big\{ \frac{\mu_{\rm d1}^{-}(Q_{{\rm det}|Z,Z, {\rm d2}}-p_{\rm t})e^{\mu_{\rm d2}^{-}}-\mu_{\rm d2}^{+}Q_{{\rm det}|Z,Z, {\rm d1}}e^{\mu_{\rm d1}^{+}}}{(1-p_{\rm t})(\mu_{\rm d1}^{-}-\mu_{\rm d2}^{+})}, 0 \Big\}.
\end{align}
In addition, $Q_{{\rm u},1,{\rm det}|Z,Z,s}$ in Eq. (\ref{keyrate}) is lower bounded by
\begin{equation}
Q_{{\rm u},1,{\rm det}|Z,Z,s}\geq Q_{{\rm u},1,{\rm det}|Z,Z,s}^{-}=(1-p_{\rm t})\mu_{\rm s}^{-}e^{-\mu_{\rm s}^{-}}Y_{{\rm det}|Z,Z,{\rm u},1}^{-},
\end{equation}
where $\mu_{\rm s}^{-}e^{-\mu_{\rm s}^{-}}$ is the lower bound on the probability of a signal pulse becoming a single-photon state.

Similarly, $Y_{b|c,X,{\rm u},1}$ can be lower bounded by a function of $Q_{b|c,X,k,{\rm u}}$ with $k\in\mathcal{K}$, which is the fraction of the number of the events where Bob has obtained $b$ given that Alice has sent the untagged states under the setting $\{c\in\mathcal{C}\land W'=X\land k\in\mathcal{K}\}$.
Similar to Eq. (\ref{decoy1}), we have the following general condition:
\begin{equation}
\label{decoy2}
\frac{Q_{b|c,X,k}-p_{\rm t}}{1-p_{\rm t}}\leq Q_{b|c,X,k,{\rm u}}\leq \frac{Q_{b|c,X,k}}{1-p_{\rm t}}.
\end{equation}
Combining the bounds given in \cite{mizutani} and Eq. (\ref{decoy2}), we obtain the lower and upper bound on $Y_{b|c,X,{\rm u},1}$ as
\begin{align}
\label{Ydetzz1b}
&Y_{b|c,X,{\rm u},1}\geq  Y_{b|c,X,{\rm u},1}^{-}=\frac{\mu_{\rm s}^{-}}{(\mu_{\rm s}^{-}-\mu_{\rm d1}^{+}-\mu_{\rm d2}^{-})(\mu_{\rm d1}^{+}-\mu_{\rm d2}^{-})} \notag \\ &\times\Big\{ \frac{(Q_{b|c,X, {\rm d1}}-p_{\rm t})e^{\mu_{\rm d1}^{-}}-Q_{b|c,X, {\rm d2}}e^{\mu_{\rm d2}^{+}}}{1-p_{\rm t}}\notag \\ & -\frac{(\mu_{\rm d1}^{+})^2-(\mu_{\rm d2}^{-})^2}{(\mu_{\rm s}^{-})^2}\Big(\frac{Q_{b|c,X, {\rm s}}e^{\mu_{\rm s}^{+}}}{1-p_{\rm t}}-Y_{b|c,X,{\rm u},0}^{-}\Big) \Big\},
\end{align}
where $Y_{b|c,X,{\rm u},0}^{-}$ is the lower bound on $Y_{b|c,X,{\rm u},0}$, which is the fraction of the number of events where Bob has  obtained $b$ given that Alice has sent the untagged vacuum states under the setting $\{c\in\mathcal{C}\land W'=X\}$. 
This is given by
\begin{align}
\label{Ydetzz0b}
&Y_{b|c,X,{\rm u},0}^{-}=\notag \\ &{\rm max}\Big\{ \frac{\mu_{\rm d1}^{-}(Q_{b|c,X, {\rm d2}}-p_{\rm t})e^{\mu_{\rm d2}^{-}}-\mu_{\rm d2}^{+}Q_{b|c,X, {\rm d1}}e^{\mu_{\rm d1}^{+}}}{(1-p_{\rm t})(\mu_{\rm d1}^{-}-\mu_{\rm d2}^{+})}, 0 \Big\}.
\end{align}
In addition, $Y_{b|c,X,{\rm u},1}$ is upper bounded by
\begin{align}
\label{YbcX1}
Y_{b|c,X,{\rm u},1}&\leq Y_{b|c,X,{\rm u},1}^{+}\notag \\
&=\frac{Q_{b|c, X, {\rm d1}}e^{\mu_{\rm d1}^{+}}-(Q_{b|c, X, {\rm d2}}-p_{\rm t})e^{\mu_{\rm d2}^{-}}}{(1-p_{\rm t})(\mu_{\rm d1}^{-}-\mu_{\rm d2}^{+})}.
\end{align}

\section{Upper bound on $\bm{Y_{a,b|{\rm vir,{\rm u}}}}$}
In this appendix, we present the derivation of $Y_{a,b|{\rm vir,{\rm u}}}$ in Eq. (\ref{shortcal}) in the main text. 
As $\hat{P}(\ket{\theta})$ is written as  $\hat{P}(\ket{\theta})=(\hat{I_d}+\sin{\theta}\hat{X}+
\cos{\theta}\hat{Z})/2$ by using Pauli operators, we obtain 
\begin{equation}
\label{relation}
  \begin{pmatrix}
      $Tr$[\hat{D}_{{b_X}}^{(i)}\hat{P}(\ket{\theta_{0_Z}^{(i)}})] \\
      $Tr$[\hat{D}_{{b_X}}^{(i)}\hat{P}(\ket{\theta_{1_Z}^{(i)}})] \\
      $Tr$[\hat{D}_{{b_X}}^{(i)}\hat{P}(\ket{\theta_{0_X}^{(i)}})]  
    \end{pmatrix} = 
    M^{(i)}
 \begin{pmatrix}
      $Tr$[\hat{D}_{{b_X}}^{(i)} \hat{I_d}/2] \\
      $Tr$[\hat{D}_{{b_X}}^{(i)} \hat{X}/2] \\
      $Tr$[\hat{D}_{{b_X}}^{(i)} \hat{Z}/2] 
    \end{pmatrix},
\end{equation}
where $M^{(i)}:=(\vec{V}_{0_Z}, \vec{V}_{1_Z}, \vec{V}_{0_X})^{T}$ with $\vec{V}_{c}^{T}:=(1, \sin{\theta_{c}^{(i)}}, \cos{\theta_{c}^{(i)}})$, with  $T$ representing the transpose. 
Since the Bloch vectors of $\hat{P}(\ket{\theta_{0_Z}^{(i)}})$, $\hat{P}(\ket{\theta_{1_Z}^{(i)}})$, and $\hat{P}(\ket{\theta_{0_X}^{(i)}})$ form a triangle (from the assumption A-3), ${\rm rank}M^{(i)}=3$ is satisfied, which means there always exists the inverse of $M^{(i)}$. Therefore, according to Eq. (\ref{relation}), we can obtain the transmission rate of Pauli matrices ${\rm Tr}[\hat{D}_{{b_X}}^{(i)} \hat{I_d}/2]$, ${\rm Tr}[\hat{D}_{{b_X}}^{(i)} \hat{X}/2]$, and ${\rm Tr}[\hat{D}_{{b_X}}^{(i)} \hat{Z}/2]$ from the ones of actually sending states. 
Combining this fact with Eqs. (\ref{azuma1}) and  (\ref{azuma2}), we have
\begin{align} 
\label{Ymax}
&Y_{a,b|{\rm vir},{\rm u}}\leq p_{a_{\rm vir}}^{+}\frac{1}{N_{{\rm s},{\rm u},1}}\sum_{i=1}^{N_{{\rm s},{\rm u},1}}\left<{\rm Tr}\left[\hat{D}_{b_X}^{(i)}\hat{P}(\ket{\theta_{a_{\rm vir}}^{(i)}}) \right]\right> \notag \\
&=p_{a_{\rm vir}}^{+}\frac{1}{N_{{\rm s},{\rm u},1}}\sum_{i=1}^{N_{{\rm s},{\rm u},1}}
\Big<{\rm Tr}\left[\hat{D}_{b_X}^{(i)}\hat{I_d}/2 \right]  \notag \\
&+\sin{\theta_{a_{\rm vir}}^{(i)}}{\rm Tr}\left[\hat{D}_{b_X}^{(i)}\hat{X}/2 \right]+\cos{\theta_{a_{\rm vir}}^{(i)}}{\rm Tr}\left[\hat{D}_{b_X}^{(i)}\hat{Z}/2 \right]\Big> \notag \\
&=p_{a_{\rm vir}}^{+}\frac{1}{N_{{\rm s},{\rm u},1}}\sum_{i=1}^{N_{{\rm s},{\rm u},1}}\sum_{c\in\mathcal{C}}\left< g_{a, c}^{(i)}{\rm Tr}\left[\hat{D}_{b_X}^{(i)}\hat{P}(\ket{\theta_{c}^{(i)}}) \right]\right>
\notag \\
&\leq p_{a_{\rm vir}}^{+}\frac{1}{N_{{\rm s},{\rm u},1}}\sum_{i=1}^{N_{{\rm s},{\rm u},1}}\sum_{c\in\mathcal{C}}g_{a, c}^{+}\left<{\rm Tr}\left[\hat{D}_{b_X}^{(i)}\hat{P}(\ket{\theta_{c}^{(i)}}) \right]\right> \notag \\
&=p_{a_{\rm vir}}^{+}\sum_{c\in\mathcal{C}}g_{a, c}^{+}Y_{b|c,X,{\rm u},1},
\end{align}
where $p_{a_{\rm vir}}^{+}$ is the maximum value of $p_{a_{\rm vir}}^{(i)}$ under the constraint of $\theta_{c}^{-}\leq \theta_{c}^{(i)} \leq \theta_{c}^{+}$.
Here, $g_{a, c}^{(i)}$ and $g_{a, c}^{+}$  with $a\in\{0,1\}$ and  $c\in\mathcal{C}$ in Eq. (\ref{Ymax}) is a function of $\theta_{0_Z}^{(i)}$, $\theta_{1_Z}^{(i)}$, and $\theta_{0_X}^{(i)}$, and its upper bound under the condition of $\theta_{c}^{-}\leq \theta_{c}^{(i)} \leq \theta_{c}^{+}$, respectively. $g_{a, c}^{(i)}$ is a parameter that connects the transmission rate of Pauli matrices to that of actual states.

In what follows, we show the derivation of $p_{a_{\rm vir}}^{+}$ for all $a\in\{0,1\}$, the explicit formula of $g_{a, c}^{(i)}$, and the way to derive $g_{a, c}^{+}$ for all  $a\in\{0,1\}$ and $c\in\mathcal{C}$. 
The derivation of these upper bounds depends on the range $[\theta_{c}^{-}, \theta_{c}^{+}]$, which Alice identifies during the experiments. Here, as an example, we consider a special case that we consider in our simulation. 
That is, the following conditions are satisfied:
\begin{align}
\label{simulationcondition}
-\theta\leq\theta_{0_Z}^{(i)}&\leq\theta, \notag \\
\pi-\theta\leq\theta_{1_Z}^{(i)}&\leq\pi+\theta, \notag \\
\pi/2-\theta\leq\theta_{0_X}^{(i)}&\leq\pi/2+\theta.
\end{align}
In the following, we begin with deriving $p_{a_{\rm vir}}^{+}$. Then we show the way to obtain $g_{a,c}^{+}$. Note that, $\theta$, which reflects the accuracy of the phase modulator, is less than  $10^\circ$ in our simulation, and we use this fact in the following calculation.
Also, note that our method can be generalized for any $[\theta_{c}^{-}, \theta_{c}^{+}]$. 

{\it \textbf{Computation of $\bm{p_{a_{\rm vir}}^{+}$}}} --
We apply Eq. (\ref{simulationcondition}) to Eq. (\ref{VirPro}), obtaining
\begin{align}
p_{0_{\rm vir}}^{(i)}&=\frac{1}{2}\left(1+\cos{\left(\frac{\theta_{0_Z}^{(i)}-\theta_{1_Z}^{(i)}}{2}\right)}\right) \notag \\
&\leq\frac{1}{2}(1+\sin\theta)=p_{0_{\rm vir}}^{+},
\end{align}
and
\begin{align}
p_{1_{\rm vir}}^{(i)}&=\frac{1}{2}\left(1-\cos{\left(\frac{\theta_{0_Z}^{(i)}-\theta_{1_Z}^{(i)}}{2}\right)}\right) \notag \\
&\leq\frac{1}{2}(1+\sin\theta)=p_{1_{\rm vir}}^{+},
\end{align}
where we have used the fact that $-\pi/2-\theta\leq(\theta_{0_Z}^{(i)}-\theta_{1_Z}^{(i)})/2\leq-\pi/2+\theta$
.

{\it \textbf{Computation of  $\bm{g_{a,c}^{+}$}}} --
$g_{a,c}^{(i)}$ with $a\in\{0,1\}$ and $c\in\mathcal{C}$ can be written as a function of $u_{a, c}^{(i)}(\theta_{0_Z}^{(i)},\theta_{1_Z}^{(i)},\theta_{0_X}^{(i)})$ and $v_{a, c}^{(i)}(\theta_{0_Z}^{(i)},\theta_{1_Z}^{(i)},\theta_{0_X}^{(i)})$, which are defined later. 
First, we find the range of $u_{a, c}^{(i)}$ and $v_{a, c}^{(i)}$ under  Eq. (\ref{simulationcondition}). Then, we compute the partial derivative of  $g_{a,c}^{(i)}$ with respect to $u_{a, c}^{(i)}$ and $v_{a, c}^{(i)}$, and check whether the signs of these partial derivatives are positive or not. Finally, by using these signs, we identify the values of $u_{a, c}^{(i)}$ and $v_{a, c}^{(i)}$ that maximize $g_{a,c}^{(i)}$ to obtain $g_{a,c}^{+}$.
Note that this upper bound is not the maximum value of $g_{a,c}^{(i)}$ under  Eq. (\ref{simulationcondition}) because we  regard $u_{a,c}^{(i)}$ and $v_{a,c}^{(i)}$ as individual variables although this is not the case. Indeed, the upper bound is certainly greater than or equal to the maximum value. Therefore, we successfully estimate the upper bound on $g_{a,c}^{(i)}$. In the following, we concisely summarize the formulas of $g_{a,c}^{(i)}$, and the upper bounds:
\begin{equation}
\bullet \ g_{0,0_Z}^{(i)}=\frac{u_{0,0_Z}^{(i)}}{u_{0,0_Z}^{(i)}-v_{0,0_Z}^{(i)}}\leq\frac{\sin{\theta}}{\sin{\theta}+\cos{\frac{3}{2}\theta}}=g_{0,0_Z}^{+},
\end{equation}
where $u_{0,0_Z}^{(i)}:=\sin{((\theta_{0_Z}^{(i)}+\theta_{1_Z}^{(i)}
-2\theta_{0_X}^{(i)})/4)}$
 and $v_{0,0_Z}^{(i)}:=\sin{((-3\theta_{0_Z}^{(i)}+\theta_{1_Z}^{(i)}
+2\theta_{0_X}^{(i)})/4)}$. Due to Eq. (\ref{simulationcondition}), we have $-\sin\theta\leq u_{0,0_Z}^{(i)}\leq\sin\theta$ and $\cos\frac{3}{2}\theta\leq v_{0,0_Z}^{(i)}\leq1$. 
\begin{equation}
\bullet \ g_{0,1_Z}^{(i)}=\frac{u_{0,1_Z}^{(i)}}{u_{0,1_Z}^{(i)}+v_{0,1_Z}^{(i)}}\leq\frac{\sin{\theta}}{\sin{\theta}+\cos{\frac{3}{2}\theta}}=g_{0,1_Z}^{+},
\end{equation}
where $u_{0,1_Z}^{(i)}:=\sin{((\theta_{0_Z}^{(i)}+\theta_{1_Z}^{(i)}
-2\theta_{0_X}^{(i)})/4)}$
 and $v_{0,1_Z}^{(i)}:=\sin{((-\theta_{0_Z}^{(i)}+3\theta_{1_Z}^{(i)}
-2\theta_{0_X}^{(i)})/4)}$. Due to Eq. (\ref{simulationcondition}), we have $-\sin\theta\leq u_{0,1_Z}^{(i)}\leq\sin\theta$ and $\cos\frac{3}{2}\theta\leq v_{0,1_Z}^{(i)}\leq1$.
\begin{equation}
\bullet \ g_{0,0_X}^{(i)}=\frac{1-u_{0,0_X}^{(i)}}{v_{0,0_X}^{(i)}-u_{0,0_X}^{(i)}}\leq\frac{1-\sin{\theta}}{\cos{2\theta}-\sin{\theta}}=g_{0,0_X}^{+},
\end{equation}
where $u_{0,0_X}^{(i)}:=\cos{((\theta_{0_Z}^{(i)}-\theta_{1_Z}^{(i)})/2)}$
 and $v_{0,0_X}^{(i)}:=\cos{((\theta_{0_Z}^{(i)}+\theta_{1_Z}^{(i)}
-2\theta_{0_X}^{(i)})/2)}$. Due to Eq. (\ref{simulationcondition}), we have $-\sin\theta\leq u_{0,0_X}^{(i)}\leq\sin\theta$ and $\cos{2\theta}\leq v_{0,0_X}^{(i)}\leq1$.
\begin{equation}
\bullet \ g_{1,0_Z}^{(i)}=\frac{u_{1,0_Z}^{(i)}}{u_{1,0_Z}^{(i)}-v_{1,0_Z}^{(i)}}\leq\frac{\cos{\theta}}{\cos{\theta}-\sin{\frac{3}{2}\theta}}=g_{1,0_Z}^{+},
\end{equation}
where $u_{1,0_Z}^{(i)}:=\cos{((\theta_{0_Z}^{(i)}+\theta_{1_Z}^{(i)}
-2\theta_{0_X}^{(i)})/4)}$
 and $v_{1,0_Z}^{(i)}:=\cos{((-3\theta_{0_Z}^{(i)}+\theta_{1_Z}^{(i)}
+2\theta_{0_X}^{(i)})/4)}$. Due to Eq. (\ref{simulationcondition}), we have $\cos{\theta}\leq u_{1,0_Z}^{(i)}\leq1$ and $-\sin\frac{3}{2}\theta\leq v_{1,0_Z}^{(i)}\leq\sin\frac{3}{2}\theta$. 
\begin{equation}
\bullet \ g_{1,1_Z}^{(i)}=\frac{u_{1,1_Z}^{(i)}}{u_{1,1_Z}^{(i)}-v_{1,1_Z}^{(i)}}\leq\frac{\cos{\theta}}{\cos{\theta}-\sin{\frac{3}{2}\theta}}=g_{1,1_Z}^{+},
\end{equation}
where $u_{1,1_Z}^{(i)}:=\cos{((\theta_{0_Z}^{(i)}+\theta_{1_Z}^{(i)}
-2\theta_{0_X}^{(i)})/4)}$
 and $v_{1,1_Z}^{(i)}:=\cos{((-\theta_{0_Z}^{(i)}+3\theta_{1_Z}^{(i)}
-2\theta_{0_X}^{(i)})/4)}$. Due to Eq. (\ref{simulationcondition}), we have $\cos{\theta}\leq u_{1,1_Z}^{(i)}\leq1$ and $-\sin\frac{3}{2}\theta\leq v_{1,1_Z}^{(i)}\leq\sin\frac{3}{2}\theta$.
\begin{equation}
\bullet \ g_{1,0_X}^{(i)}=\frac{-1-u_{1,0_X}^{(i)}}{v_{1,0_X}^{(i)}-u_{1,0_X}^{(i)}}\leq-\frac{1-\sin{\theta}}{1+\sin{\theta}}=g_{1,0_X}^{+},
\end{equation}
where $u_{1,0_X}^{(i)}:=\cos{((\theta_{0_Z}^{(i)}-\theta_{1_Z}^{(i)}
)/2)}$
 and $v_{1,0_X}^{(i)}:=\cos{((\theta_{0_Z}^{(i)}+\theta_{1_Z}^{(i)}
-2\theta_{0_X}^{(i)})/2)}$. Due to Eq. (\ref{simulationcondition}), we have $-\sin\theta\leq u_{1,0_X}^{(i)}\leq\sin\theta$ and $\cos{2\theta}\leq v_{1,0_X}^{(i)}\leq1$.

\section{Detail of the simulation}

In this appendix, we present the calculations used to obtain FIGs. \ref{s1}  and \ref{s2} in the main text. Specifically, we show how to simulate the parameters that are measured in the experiments. In both simulations, 
$\theta_{c}^{(1)},\ldots,\theta_{c}^{(N)}$ are I.I.D. and follow the same Gaussian distribution which has a mean of $\bar{\theta}_{c}$ and a standard deviation of $\theta/5.33$, so that $\delta_{\theta}=10^{-7}$ is satisfied. 
Here, $\bar{\theta}_{0_Z}=0$, $\bar{\theta}_{1_Z}=\pi$, and $\bar{\theta}_{0_X}=\pi/2$. In this case, the sending single-photon state is either one of $\hat{\rho}_{0_Z}:=(\hat{I_d}+r\hat{Z})/2$, $\hat{\rho}_{1_Z}:=(\hat{I_d}-r\hat{Z})/2$, and $\hat{\rho}_{0_X}:=(\hat{I_d}+r\hat{X})/2$, where $r=\exp{\{-(\theta/5.33)^2/2)\}}$, depending on the bit and basis setting $c\in\mathcal{C}$. Note that these I.I.D. properties are not needed for the security proof, but they are assumed to simulate  the experimentally available data.

In the first simulation, we assume that Alice has a single photon source in order to see the effect of the PM error. In this case, the intensity of each pulse does not fluctuate and $\delta_{\mu}=0$, which leads to the tagging probability of $p_{\rm t}=\delta_{\theta}$. 
The fraction of the number of events where Bob obtains $b\in\{0,1\}$ given the setting $\{c\in\mathcal{C},W'\in\{Z, X\}\}$, $Q_{b|c, W'}$, is given by
\begin{align}
\label{sps1}
&Q_{b|c, W'}=\eta{\rm Pr}\left[b|\hat{\rho}_c, W'\right](1-p_{\rm d})\notag \\&+(1-\eta)p_{\rm d}(1-p_{\rm d}) +\frac{1}{2}\left\{ \eta p_{\rm d}+(1-\eta)p_{\rm d}^2 \right\},
\end{align}
where $\eta$ denotes the overall transmittance, ${\rm Pr}\left[b|\hat{\rho}_c, W'\right]:={\rm Tr}[\hat{P}(\ket{b_{W'}})\hat{\rho}_c]$ is the conditional probability that Bob obtains $b\in\{0,1\}$ given that he measures $\hat{\rho}_c$ in the $W'$ basis, and $p_{\rm d}$ is the dark count rate of Bob's detector. Here, $\eta$ is written as $\eta=\eta_{\rm ch}\eta_{\rm B}/2$, which is the product of the channel transmittance $\eta_{\rm ch}$, the detection efficiency of Bob's device $\eta_{\rm B}$, and $1/2$ of the efficiency of Bob's Mach-Zehnder interferometer. The first (second) term in Eq. (\ref{sps1}) models a single click at Bob's side produced by a photon (dark count), while the last term represents the simultaneous clicks. Note that in this last case (simultaneous clicks), Bob assigns a random bit value to the measurement outcome. 
The gain in the $Z$ basis $Q_{{\rm det}|Z, Z}$ is given by
\begin{equation}
Q_{{\rm det}|Z, Z}=1-(1-p_{\rm d})^2(1-\eta),
\end{equation}
and the untagged-photon part $Q_{{\rm u, det}|Z, Z}$ is lower bounded by
\begin{equation}
Q_{{\rm u, det}|Z, Z}\geq Q_{{\rm u, det}|Z, Z}^{-}=Q_{{\rm det}|Z, Z}-p_{\rm t}.
\end{equation}
The upper bound on the phase error rate that originates from the untagged states $e_{X,{\rm u}}^{+}$ is calculated from the following parameters:
\begin{equation}
Y_{{\rm det}|Z, Z, {\rm u}}\geq Y_{{\rm det}|Z, Z, {\rm u}}^{-}=\frac{Q_{{\rm u, det}|Z, Z}^{-}}{1-p_{\rm t}},
\end{equation}
and
\begin{equation}
\frac{Q_{b|c,X}-p_{\rm t}}{1-p_{\rm t}} \leq Y_{b|c,X, {\rm u}}\leq \frac{Q_{b|c,X}}{1-p_{\rm t}}.
\end{equation}
In addition, the bit error rate $e_Z$ is given by
\begin{equation}
e_Z=\frac{1}{Q_{{\rm det}|Z, Z}}\left(\frac{1}{2}Q_{1|0_Z, Z}+\frac{1}{2}Q_{0|1_Z, Z} \right).
\end{equation}
Finally, the asymptotic key generation rate $R$ (per pulse) is lower bounded by
\begin{equation}
R\geq p_{Z}^2\left\{Q_{{\rm u, det}|Z, Z}^{-}\left(1-h(e_{X,{\rm u}}^{+})\right)-Q_{{\rm det|Z,Z}}f_{\rm EC}h(e_Z)\right\},
\end{equation}
and the result is shown in FIG. \ref{s1} in the main text.

In the second simulation, Alice is assumed to have a coherent source with both the PM and AM errors.
The gain in the $Z$ basis is given by
\begin{align}
Q_{{\rm det}|Z,Z,k}=\int_{-\infty}^{\infty}f_G(\mu|\bar{\mu}_k, \sigma_k)\left\{ 1-(1-p_{\rm d})^2e^{-\mu\eta} \right\}d\mu,
\end{align}
where $f_G(\mu|\bar{\mu}_k, \sigma_k)$ is the Gaussian distribution function which has the mean $\bar{\mu}_k$ and the standard deviation  $\sigma_k=(\mu_{k}^{+}-\bar{\mu}_{k})/5.33$, and $\eta$ is the overall transmittance for a single photon. Here, $\exp{\left(-\mu\eta\right)}$ is the overall probability that a coherent light with intensity $\mu$ is completely lost in the channel. 
Next, let $p_{{b,{\rm click}}|c,W',\mu}$ be the conditional probability that the detector corresponding to the outcome $b\in\{0,1\}$ clicks given that the intensity of coherent state has been $\mu$ under the setting $\{c\in\mathcal{C}\land W'\in\{Z, X\}\}$, which is given by
\begin{equation}
p_{{b,{\rm click}}|c,W',\mu}=p_{\rm d}+(1-p_{\rm d})\left\{ 1-\exp\left(-{\rm Pr}\left[b|\hat{\rho}_c, W'\right]\mu\eta\right) \right\}.
\end{equation}
Therefore, the conditional probability that Bob obtains $b$ given the setting $\{c\in\mathcal{C}\land W'\in\{Z, X\}\land k\in\mathcal{K}\}$ is given by
\begin{align}
&Q_{b|c,W',k}\notag \\ &=\int_{-\infty}^{\infty}f_G(\mu|\bar{\mu}_k, \sigma_k)\Big\{ p_{{b,{\rm click}}|c,W',\mu}(1-p_{{\bar{b},{\rm click}}|c,W',\mu})\notag \\&+\frac{1}{2}p_{{b,{\rm click}}|c,W',\mu}p_{{\bar{b},{\rm click}}|c,W',\mu} \Big\} d\mu,
\end{align}
where $\bar{b}=1 (0)$ when $b=0 (1)$.
In addition, the bit error rate $e_Z$ is given by
\begin{equation}
e_Z=\frac{1}{Q_{{\rm det}|Z, Z, {\rm s}}}\left(\frac{1}{2}Q_{1|0_Z, Z, {\rm s}}+\frac{1}{2}Q_{0|1_Z, Z, {\rm s}} \right).
\end{equation}

\end{document}